%% file: combined.tex
\documentclass[prx,aps,amsfonts,amsmath,amssymb,twocolumn,superscriptaddress]{revtex4-2}

\usepackage{graphicx}
\usepackage{dcolumn}
\usepackage{appendix}
\usepackage{amsmath}
\usepackage{braket}
\usepackage{color}
\usepackage{bm}
\usepackage{multirow}
\usepackage{float}
\usepackage{natbib}

\usepackage{chapterbib}

\begin{document}

\makeatletter
\let\auto@bib\@empty
\let\auto@bib@innerbib\@empty
\let\write@bibliographystyle\relax
\makeatother

\include{arxiv_main}

\include{arxiv_si}

\end{document}

%% file: arxiv_si.tex
\onecolumngrid

\begin{center}
{\large\bfseries
Supplementary Information\\
Energy- and momentum-resolved Cooper-pair emission
from a cuprate superconductor}
\end{center}

\setcounter{section}{0}
\setcounter{subsection}{0}
\setcounter{figure}{0}
\setcounter{table}{0}
\setcounter{equation}{0}

\renewcommand{\thefigure}{S\arabic{figure}}
\renewcommand{\thetable}{S\arabic{table}}

\input{SI_body}

\makeatletter
\bibliographystyle@latex{naturemag}
\makeatother
\bibliography{bib}

%% file: SI_body.tex
\title{Supplementary Information\\ Energy- and momentum-resolved Cooper-pair emission from a cuprate superconductor}

\author{Henry Amir*, Jack Zwettler*, Ka Ho Wong, Genda Gu, Rafael M. Fernandes, Jorge Noronha, Dirk K. Morr, and Fahad Mahmood} 

\maketitle    

\setcounter{figure}{0}
\renewcommand{\figurename}{Supplementary Fig.}

\setcounter{table}{0}
\renewcommand{\tablename}{Supplementary Table}
\renewcommand{\thetable}{\arabic{table}}

\setlength{\tabcolsep}{12pt}
\renewcommand{\arraystretch}{1.5}

\newcommand{\ee}{\mathrm{e}}
\newcommand{\dd}{\mathrm{d}}
\newcommand{\SC}{\mathrm{SC}}
\newcommand{\bg}{\mathrm{bg}}
\newcommand{\pair}{\mathrm{pair}}

\newcommand{\bb}{\boldsymbol}

\section{Illustrative simulation of the Cooper-pair emission peak}

To illustrate the characteristic two-electron emission peak expected from the superconducting condensate, we calculate the pair-photoemission signal for a $d$-wave superconductor with zero pair momentum ($\mathbf{k}_1=-\mathbf{k}_2$). This calculation provides the illustrative joint energy distribution shown in Fig.~1b of the Main Text. The Bi-2212 Fermi surface is modeled using the following tight-binding dispersion
\begin{equation}
\begin{split}
\xi_{\mathbf{k}}&=-2t_1(\cos{k_x}+\cos{k_y})-4t_2 \cos{k_x}\cos{k_y}\\
&\qquad -2t_3(\cos{2k_x}+\cos{2k_y})-4t_4(\cos{2k_x}\cos{k_y}+\cos{2k_y}\cos{k_x})-\mu
\end{split}
\end{equation}
where $t_1=368~\mathrm{meV}$, $t_2=-0.342t_1$, $t_3=0.11t_1$, $t_4=-0.087t_1$ are the hopping amplitudes and $\mu=440~\mathrm{meV}$ is the chemical potential. The superconducting order parameter is given by
\begin{equation}
\Delta_{\mathbf{k}}=\frac{\Delta_0}{2}(\cos{k_x}-\cos{k_y})
\end{equation}
where $\Delta_0=30~\mathrm{meV}$. Suppose that the two photoelectrons carry momenta $\mathbf{k}_1=(k_{x},k_y)$ and $\mathbf{k}_2=(-k_{x},-k_y)$ and opposite spin, the pair photoemission rate is given by \cite{Mahmood2022}:
\begin{equation}
    P^{(2)}_{SC}(k_{x},k_{y};\Gamma)=2\pi\left[\frac{1}{\pi}\frac{\Gamma}{(\hbar\omega-2\epsilon_{\mathbf{k}})^2+\Gamma^2}\right]\Bigg\vert \int\frac{\mathrm{d}^2\mathbf{p}}{(2\pi)^2}\frac{V(\mathbf{p}-\mathbf{k})}{\hbar\omega-\epsilon_{\mathbf{p}}-E_{\mathbf{p}}+i\delta}\frac{\Delta_{\mathbf{p}}}{2E_{\mathbf{p}}}\Bigg\vert^2
\end{equation}
where $\hbar\omega$ is the photon energy, $\epsilon_{\mathbf{k}}=\frac{\hbar^2 k^2}{2m_e a_0^2}+W$ is the sum of the photoelectron's kinetic energy and work function, and $E_{\mathbf{p}}=\sqrt{\xi_{\mathbf{p}}^2+\Delta_{\mathbf{p}}^2}$ is the dispersion of the superconducting state.

To account for finite momentum broadening in the simulated pair photoemission distribution, we introduce Lorentzian broadening of the theoretical result. We define
\begin{equation}
    L(x;\gamma)=\frac{1}{\pi}\frac{\gamma}{x^2+\gamma^2}
\end{equation}
where $\gamma$ is the Lorentzian half-width at half maximum. The broadened pair photoemission rate is then
\begin{equation}
\begin{split}
    P^{(2)}_{conv}(\mathbf{k}_1,\mathbf{k}_2)&=L(k_{1x}+k_{2x};\gamma_{sum})L(k_{1y}+k_{2y};\gamma_{sum})\\
    &\qquad \times \sum_{n}L(k_{1x}-k_{2x}-2k_{x}^{(n)};\gamma_{diff})L(k_{1y}-k_{2y}-2k_{y}^{(n)};\gamma_{diff})P^{(2)}_{SC}(k_{x}^{(n)},k_{y}^{(n)};\Gamma)
\end{split}
\end{equation}
Here, the summation is performed over a discrete momentum grid within a box of width $0.2a_0^{-1}$ centered at the antinodal point of the $d$-wave Fermi surface to reduce computational cost.

We introduce distinct Lorentzian broadening widths, $\gamma_{\mathrm{sum}}$ and $\gamma_{\mathrm{diff}}$, for the sum momentum $\mathbf{k}_1+\mathbf{k}_2$ and difference momentum $\mathbf{k}_1-\mathbf{k}_2$, respectively. Allowing these widths to vary independently provides a phenomenological description of anisotropic momentum broadening in the simulated two-electron distribution.

\newpage

\section{Experimental details and single-electron ARPES characterization}

\subsection{Sample information}
Optimally doped Bi$_2$Sr$_2$CaCu$_2$O$_{8+\delta}$ (Bi-2212) single crystals, with cross-sectional areas of approximately 1--4\,mm$^2$, were mounted on copper sample plates using conductive silver epoxy to provide thermal and electrical contact with the manipulator. Samples were cleaved \emph{in situ} under ultrahigh vacuum by removing a post attached to the crystal, exposing a fresh surface parallel to the crystallographic $ab$ plane. Before acquiring ARPES or two-electron ARPES (2e-ARPES) data, we mapped the photoemission intensity over the exposed surface to select a spatially uniform measurement region. The incident beam had a spot diameter of approximately 200~$\mu$m, substantially smaller than the lateral dimensions of the crystals.

\subsection{Tight-binding dispersion and accessible momentum regions}
The measured Fermi surface (Fig.~2 of the Main Text) is consistent with the published tight-binding parametrization of Bi-2212 \cite{Bi2212_gap_sign}. For comparison with the experimental momentum coverage, we show the bonding-band dispersion and omit superstructure replicas \cite{Bi2212_gap_sign_2,Bi2212_ARPES_1}. The highlighted regions in Fig.~2 (Main Text) indicate the in-plane momenta accessible to the two detectors for the geometries summarized in Supplementary Table~\ref{table:ARPES}.

The magnitude of the photoelectron momentum parallel to the sample surface is determined from its kinetic energy and emission angle:
\begin{equation}
    k_{||} = \sqrt{\frac{2m_eE_k}{\hbar^2}}\mathrm{sin}\theta,
    \label{eq:ARPES}
\end{equation}
where $m_e$ is the electron mass, $E_k$ is the photoelectron kinetic energy, $\hbar$ is the reduced Planck constant and $\theta$ is the emission angle relative to the surface normal. A 15$^\circ$ angular field of view was assumed for each detector. For conventional ARPES, the tabulated kinetic energy corresponds to emission from the Fermi level, $E_k=h\nu-\Phi$, where $h\nu$ is the photon energy and $\Phi$ is the sample work function. For 2e-ARPES, $E_k=h\nu/2-\Phi$ denotes the nominal kinetic energy of each electron for equal energy sharing and zero pair binding energy.

\begin{table}[ht]
    \centering
                \begin{tabular}{|c|c|c|c|c|}
                    \hline
                    Region & Detector & $E_k$ & Measurement & Surface normal \\
                    \hline
                    1 & 2 & $h\nu-\Phi$ & ARPES & $\hat{x}$ \\
                    \hline
                    2 & 1 & $h\nu-\Phi$ & ARPES & $\frac{1}{\sqrt{2}}(\hat{x}-\hat{y})$ \\
                    \hline
                    3 & 1 & $h\nu-\Phi$ & ARPES & $\hat{x}$ \\
                    \hline
                    4 & 2 & $\frac{h\nu}{2}-\Phi$ & 2e-ARPES  & $\hat{x}$ \\
                    \hline
                    5 & 1 & $\frac{h\nu}{2}-\Phi$ & 2e-ARPES  & $\hat{x}$ \\
                    \hline
                \end{tabular}
    \caption{\textbf{Experimental geometries used to determine the momentum regions in Fig.~2 of the main text.} The detector axes lie along $(\hat{\mathbf{x}}-\hat{\mathbf{y}})/\sqrt{2}$ (detector 1) and $(\hat{\mathbf{x}}+\hat{\mathbf{y}})/\sqrt{2}$ (detector 2); the incident beam propagates along $-\hat{\mathbf{x}}$. All directions are specified in the laboratory coordinate system. The 2e-ARPES kinetic energies correspond to equal energy sharing at zero pair binding energy.}
    \label{table:ARPES}
\end{table}

\subsection{Gap structure of Bi-2212}

\begin{figure*}[ht]
    \centering
    \includegraphics[width=15cm]{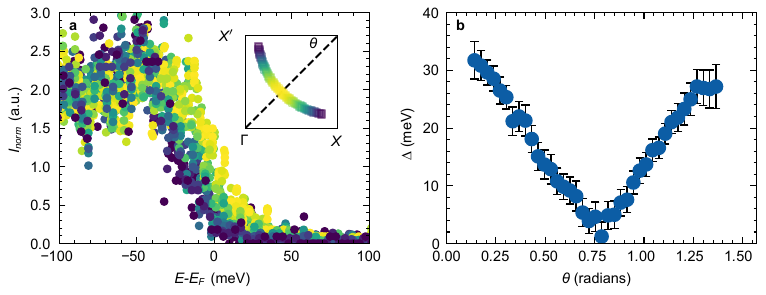}
    \caption{\textbf{Single-electron ARPES characterization of optimally doped Bi-2212 at 33~K.} \textbf{a} EDCs recorded at positions along the Fermi surface, indicated by the inset. Each EDC is normalized to its mean intensity over the energy interval $[-200,200]$~meV. \textbf{b}, Angular dependence of the gap estimate $\Delta$, determined from fits to the EDCs as described in the text. The angle $\theta$ is defined in \textbf{a}. Error bars represent 68\% confidence intervals of the fitted estimates.}
    \label{fig:S_ARPES}
\end{figure*}

The superconducting gap of our optimally-doped Bi-2212 samples was characterized by conventional ARPES. Supplementary Fig.~\ref{fig:S_ARPES}a shows energy-distribution curves (EDCs) at points along the Fermi surface, parametrized by the Fermi-surface angle $\theta$ defined in the figure. The leading-edge position shifts to higher binding energy away from the nodal direction ($\theta=\pi/4$), consistent with the opening of an anisotropic superconducting gap.

To quantify this shift, each EDC was fitted phenomenologically with a Fermi-Dirac edge multiplied by a linear spectral background and convolved with a Gaussian broadening function:
\begin{equation}
    I(E)=\left[\frac{A+BE}{1+\exp\!\left(\frac{E-\mu}{k_{\mathrm B}T}\right)}\right]
    *\frac{1}{\sqrt{2\pi}\sigma}\exp\!\left(-\frac{E^2}{2\sigma^2}\right),
    \label{eq:FD}
\end{equation}
where $E$ is electron energy relative to the experimental energy reference, $T$ is the sample temperature, $A$ and $B$ describe the locally linear prefactor, $\mu$ is the fitted edge position and $\sigma$ is the Gaussian broadening width. The symbol $*$ denotes convolution over energy. The parameter $\mu$ is used here only to track the leading-edge position; it is not interpreted as a momentum-dependent thermodynamic chemical potential. Fits based on a Dynes spectral function \cite{Dynes} were not sufficiently stable at the available energy resolution and counting statistics.

Taking the difference between $\mu$ and the Fermi level at the node, $E_F$, we obtain an estimate for the gap $\Delta$.  Its angular dependence is consistent with the expected $d$-wave gap anisotropy of optimally doped Bi-2212 (Supplementary Fig.~\ref{fig:S_ARPES}b) \cite{Bi2212_gap_1,Bi2212_gap_sign,Bi2212_gap_2}.

\newpage

\section{2e-ARPES}
\subsection{Coincidence Algorithm and Data Analysis}

Photoelectrons were recorded using two angle-resolved time-of-flight analyzers, synchronized to the laser repetition clock. Each detection event was assigned a laser-pulse index $n$, an energy $E_i$, and an in-plane momentum $\mathbf{k}_i$, where $i=1,2$ labels the detector. We collectively denote the energy and momentum coordinates by $\xi_i=(E_i,\mathbf{k}_i)$. 

A same-pulse coincidence was registered when each detector recorded exactly one electron originating from the same laser pulse. Such events can arise from two distinct processes: true double photoemission, in which a single photon induces the emission of two electrons, or accidental coincidences, in which independent single-electron photoemission events occur within the same laser pulse. Since both processes contribute to the measured coincidence distribution, the accidental background must be determined independently to isolate the correlated two-electron signal.

The measured same-pulse coincidence distribution is denoted by $I^{(2)}(\xi_1,\xi_2)$. For statistically independent single-electron emission, the expected accidental-coincidence distribution factorizes as
\begin{equation}
I_{\mathrm{acc}}^{(2)}(\xi_1,\xi_2)
=
I^{(1)}(\xi_1)I^{(1)}(\xi_2),
\end{equation}
where $I^{(1)}(\xi_i)$ is the appropriately normalized single-electron intensity in detector $i$.

Experimentally, the accidental background is estimated directly from detections associated with different laser pulses. We define the pulse-offset coincidence histogram
\begin{equation}
M_l(\xi_1,\xi_2)
=
\sum_n
c_{1,n}(\xi_1)c_{2,n-l}(\xi_2),
\label{eq:SI_offset}
\end{equation}
where $c_{i,n}(\xi_i)$ equals one when detector $i$ records exactly one electron in bin $\xi_i$ from pulse $n$, and zero otherwise. The integer $l$ specifies the pulse offset, and the sum includes only valid pulse indices.

For $l=0$, events are paired using the same laser pulse in both detectors, giving
\begin{equation}
I^{(2)}(\xi_1,\xi_2)
=
M_0(\xi_1,\xi_2).
\end{equation}
For nonzero $l$, an event from pulse $n$ in detector 1 is paired with an event from pulse $n-l$ in detector 2. Because the two electrons originate from different laser pulses, they cannot result from the same microscopic double-photoemission event. These offset distributions therefore provide an estimate of the accidental background, provided the detection statistics are stationary and independent between pulses.

For the analysis presented in the main text, we average the two nearest-neighbor pulse offsets:
\begin{equation}
I_{\mathrm{acc}}^{(2)}(\xi_1,\xi_2)
=
\frac{1}{2}
\left[
M_{+1}(\xi_1,\xi_2)
+
M_{-1}(\xi_1,\xi_2)
\right].
\label{eq:SI_background}
\end{equation}
The difference
$I^{(2)}(\xi_1,\xi_2)-I_{\mathrm{acc}}^{(2)}(\xi_1,\xi_2)$
then gives the measured coincidence excess above the accidental background.

\begin{figure*}[ht]
    \centering
    \includegraphics[width=5.5cm]{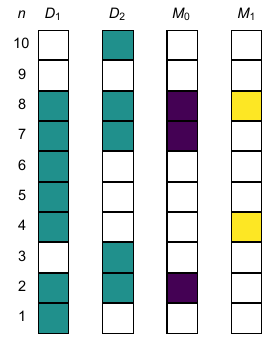}
    \caption{\textbf{Pulse-matching algorithm for 2e-ARPES.} For this generated data, detectors 1 and 2 ($D_1$ and $D_2$) are filled in for pulse $n$ if there was a detected event. $M_0$ is found by identifying the pulses labeled $n$ where both detectors observed an event. $M_{+1}$ is found by identifying cases where there was an event from pulse $n$ in detector 1 and event from pulse $n-1$ in detector 2.}
    \label{fig:algorithm}
\end{figure*}

Supplementary Fig.~\ref{fig:algorithm} illustrates the pulse-matching procedure. The synchronous histogram $M_0$ is constructed by identifying pulses for which both detectors recorded an electron, whereas $M_{+1}$ is obtained by pairing detections from pulse $n$ in detector 1 with those from pulse $n-1$ in detector 2.

As in the Methods section of the main text, we characterize the correlated two-electron emission using
\begin{equation}
P^{(2)}(\xi_1,\xi_2)
=
\frac{
\left[
I^{(2)}(\xi_1,\xi_2)
-
I_{\mathrm{acc}}^{(2)}(\xi_1,\xi_2)
\right]^2
}{
I^{(2)}(\xi_1,\xi_2)
+
I_{\mathrm{acc}}^{(2)}(\xi_1,\xi_2)
+
\epsilon
},
\label{eq:SI_P2}
\end{equation}
where $\epsilon>0$ is a small regularization parameter that prevents numerical instabilities in bins containing very few counts. The numerator measures the squared excess of same-pulse coincidences over the accidental background, while the denominator weights this excess by the total counting intensity. 

\begin{figure*}[ht]
    \centering
    \includegraphics[width=8cm]{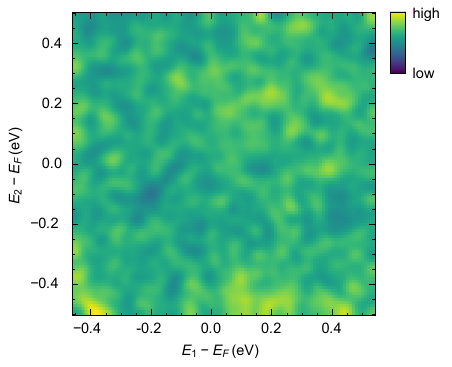}
    \caption{\textbf{2e-ARPES background.} Intensity plot of $I_B(\xi_1,\xi_2) = \sum_{\substack{l=-3\\l\neq0}}^{3} M_l(\xi_1,\xi_2)$ for the 8.5~K dataset in the main text. Intensity modulations can be attributed to shot noise.}
    \label{fig:background}
\end{figure*}

To assess the accidental-coincidence background independently, we also examine the combined histogram obtained from six nonzero pulse offsets,
\begin{equation}
I_B(\xi_1,\xi_2) = \sum_{\substack{l=-3\\l\neq0}}^{3} M_l(\xi_1,\xi_2).
\label{eq:SI_IB}
\end{equation}
Supplementary Fig.~\ref{fig:background} shows this distribution for the 8.5 K dataset used in the main text. The background exhibits no discernible structure beyond statistical counting fluctuations.

\subsection{Optimization of $P^{(2)}$}
\label{subsec:Signal-to-noise ratio}

As established in previous sections, the sensitivity of 2e-ARPES measurements is determined by the number of correlated two-electron events relative to the statistical fluctuations of the accidental-coincidence background. Consequently, the value of $P^{(2)}$ depends both on sample-dependent photoemission probabilities and on experimental parameters. The role of the latter must be characterized in order to maximize $P^{(2)}$. 

We model the detection statistics using three probabilities: $P_C$, the probability of detecting a correlated two-electron photoemission event per incident photon; $P_S$, the probability of detecting a single-electron photoemission event in either analyzer per incident photon; and $P_B$, the probability of a detector dark count within one laser-pulse window. We assume comparable detection probabilities for the two analyzers and $P_C\ll P_S$. Additionally, $P_C$ and $P_S$ are inherently functions of $\xi_1$ and $\xi_2$.

For a Poisson distribution of incident photons with mean $\lambda$ per laser pulse, the probability of detecting a true two-electron event is, in the low-count-rate limit,
\begin{equation}
P_{\mathrm{true}}^0 \simeq \lambda P_C.
\end{equation}
Accidental coincidences arise when the two analyzers independently detect single photoelectrons or dark counts within the same laser pulse. Their probability is approximately
\begin{equation}
P_{\mathrm{acc}}
\simeq
(\lambda P_S+P_B)^2.
\end{equation}
Here, $P_B$ scales inversely with the laser repetition rate $f$ for a fixed detector dark-count rate. These expressions apply in the regime of low detector occupancy, where multiple detections within a single laser pulse are negligible.

The finite lifetime of the freshly cleaved sample surface introduces an additional constraint on the acquisition time. We model the decrease in the correlated photoemission probability as
\begin{equation}
P_{\mathrm{true}}(t)
=
P_{\mathrm{true}}^0 e^{-t/\tau},
\end{equation}
where $t$ is the elapsed acquisition time and $\tau$ is the characteristic decay time, which depends on the sample and vacuum conditions. The single-electron background is taken to remain approximately constant over the acquisition.

For an acquisition of duration $t$, the expected number of correlated events is therefore
\begin{equation}
N_{\mathrm{true}}(t)
=
P_{\mathrm{true}}^0 f\tau
\left(1-e^{-t/\tau}\right),
\end{equation}
while the expected number of accidental coincidences is
\begin{equation}
N_{\mathrm{acc}}(t)
=
P_{\mathrm{acc}}ft.
\end{equation}

Assuming Poisson counting statistics, the variance of the background-subtracted coincidence signal is
\begin{equation}
\sigma^2
=
N_{\mathrm{true}}
+
N_{\mathrm{acc}}
\left(1+\frac{1}{\gamma}\right),
\end{equation}
where $\gamma$ is the number of independent nonzero pulse offsets used to estimate $I_{\mathrm{acc}}^{(2)}$. The factor $1+1/\gamma$ accounts for fluctuations in both the same-pulse coincidence distribution $I^{(2)}$ and the estimated accidental background. For the nearest-neighbor subtraction used in this work, $\gamma=2$.

Combining these expressions gives
\begin{equation}
P^{(2)}(t)
=
\frac{
\left[
P_{\mathrm{true}}^0 f\tau
(1-e^{-t/\tau})
\right]^2
}{
P_{\mathrm{true}}^0 f\tau
(1-e^{-t/\tau})
+
P_{\mathrm{acc}}
\left(1+\frac{1}{\gamma}\right)ft
}.
\label{eq:SI_SNR}
\end{equation}
In the limit of low photon occupancy and negligible sample degradation, this expression reduces to the squared signal-to-noise ratio discussed in Ref.~\cite{Leitner2021}.

\begin{figure*}[ht]
    \centering
    \includegraphics[width=7.5cm]{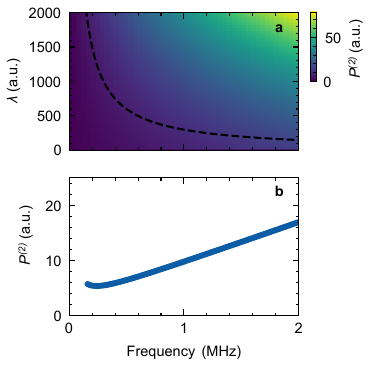}
    \caption{\textbf{Simulated $P^{(2)}$.} \textbf{a} $P^{(2)}$ integrated over $\xi_1$ and $\xi_2$ as a function of the average number of photons per pulse $\lambda$ and the laser repetition rate with $P_C=3\times10^{-8}$, $P_S = 1\times 10^{-3}$, and $\tau=3\times10^4$ s. For each pixel, the acquisition time $t$ is chosen such that it maximizes $P^{(2)}$, which is our procedure after a measurement. Due to experimental constraints, the measurement must be performed to the left of the black-dashed line. The value of $P^{(2)}$ along this line is given in \textbf{b}.}
    \label{fig:S_SNR}
\end{figure*}

Supplementary Fig.~\ref{fig:S_SNR} shows the calculated $P^{(2)}$ as a function of the laser repetition rate $f$ and the mean photon number per pulse $\lambda$, using representative experimental parameters. For each combination of $f$ and $\lambda$, the acquisition time is chosen to maximize $P^{(2)}$. The calculations illustrate the advantage of increasing the repetition rate while maintaining a fixed average photon flux. Under these conditions, the correlated event rate remains approximately constant, whereas the probability of accidental coincidences decreases because fewer photoelectrons are generated per pulse.

In practice, the accessible photon flux is limited by the detector count rate, while the available extreme-ultraviolet photon flux depends on the efficiency of high-harmonic generation. The repetition rate of $f=525$ kHz used in the main-text measurements represents a compromise between these constraints, as the high-harmonic photon flux decreases sharply at higher repetition rates in the present setup.

\subsection{Pulse independence}
\begin{figure*}[ht]
    \centering
    \includegraphics[width=14.5cm]{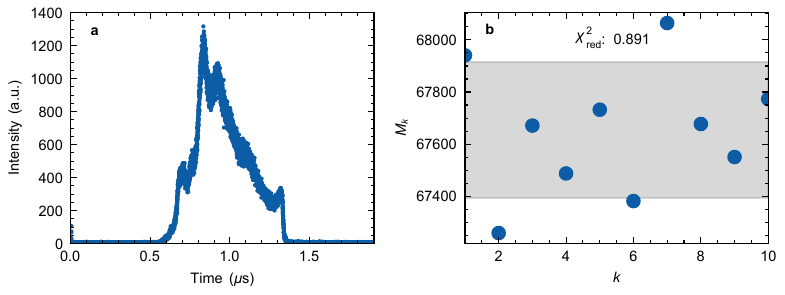}
     \caption{\textbf{Pulse independence.} \textbf{a} Histogram of detected electron time-of-flights. The x-axis span one laser period for the repetition rate $f=525$ kHz used in the experiment. \textbf{b} Coincidence counts $M_l$ as a function of pulse offset $l$ (as defined in the Methods section in the main text). The grey region encapsulates $2\sigma$, where $\sigma = \sqrt{N}$. This is representative of Poisson counting statistics, and $N$ is the average of all data points. The reduced $\chi^2$, calculated as $\chi^2_{\mathrm{red}} = \frac{1}{N(p-1)}\sum_{l =1}^p(M_l-N)^2$ where $p$ is the number of offsets, is approximately equal to 1.}
    \label{fig:S_PI}
\end{figure*}

The pulse-offset coincidence analysis requires that detections associated with successive laser pulses are statistically independent. One potential source of ambiguity arises from the spread in electron time-of-flights. Slow photoelectrons emitted by one laser pulse could overlap with faster photoelectrons emitted by a subsequent pulse. Supplementary Fig.~\ref{fig:S_PI}a shows that the detected electron time-of-flight window is shorter than the laser repetition period, $t_{\nu}=1/f$, where $f=525$ kHz is the repetition rate used in the experiment. The electronic dead time following each detection is approximately 150 ns, also substantially shorter than the laser repetition period.

A second consideration is the possibility of systematic pulse-to-pulse intensity variations. The laser system employs two Pockels cells for pulse amplification and pulse selection, respectively. The pulse picker reduces the repetition rate by transmitting every $m$th pulse, where $m\geq1$ is an integer, while maintaining approximately constant pulse energy. However, pulse selection can introduce periodic intensity modulations, potentially producing correlations between detections associated with different laser pulses.

To test for such effects, we examine the coincidence counts $M_l$ as a function of the nonzero pulse offset $l$, as defined in the preceding section. Here, $M_l$ denotes the total coincidence counts integrated over the measured energy and momentum regions. As shown in Supplementary Fig.~\ref{fig:S_PI}b, the counts for $l>0$ are consistent with a constant mean and Poisson counting fluctuations, with a reduced chi-squared value of $\chi^2_{\mathrm{red}}=0.891$. This agreement supports the assumption of statistically independent pulse-to-pulse detections underlying the accidental-coincidence background subtraction.
    
\subsection{Figure generation}
    Figures were generated using Python. Data processing, visualization, and figure formatting were performed using the python package Matplotlib and other standard Python packages and libraries, as appropriate. Constant intensity contour plots of $P^{(2)}$ were made using Matplotlib's built in Contour method, where lines are drawn connecting points of equal intensity. $P^{(2)}$ was calculated using the same method as outlined in the Coincidence Algorithm and Data Analysis section of this supplement and normalized by the total intensity. Contour lines of matching color in each plot represent the same relative intensity of $P^{(2)}$. The data used to generate the figures are available in the Supplementary Information Source Data files.

\subsection{Bootstrapping}
Statistical uncertainties in the two-electron correlation measure $P^{(2)}$ were determined using bootstrap resampling \cite{Bootstrap_intro_new}. The same-pulse and pulse-offset coincidence samples were independently resampled with replacement. For each bootstrap realization, the distributions $I^{(2)}(\xi_1,\xi_2)$ and $I_{\mathrm{acc}}^{(2)}(\xi_1,\xi_2)$ were reconstructed, and the corresponding $P^{(2)}(\xi_1,\xi_2)$ was calculated using Eq.~\ref{eq:SI_P2}. Repeating this procedure generates a distribution of $P^{(2)}$ values for each energy- and momentum-resolved bin.
    
It is standard to represent the error of a bootstrapped quantity $x$ as the product of the standard error and a critical $T$ value \cite{Bootstrap_intro_new}. For example, a critical $T$ value of 1 is associated with a 68 percent confidence interval for normally distributed data points. It follows that the lower error bar for each data point should go to the 16th percentile and the upper error bar should go to the 84th percentile of the distribution of bootstrapped means. To find these values for a discrete set of data, we can use the expression from \cite{Quantile}
\begin{equation}
   \hat{Q} = (1-g)x_j+gx_{j+1},
\end{equation}
where $j=\left[qn\right]$ and $g = qn-j$. Here, the $[x]$ operator returns the largest integer less than $x$. Also, $N$ is the number of points to average over, $n=N-1$, and $q$ is the percentile divided by 100. For example, to obtain the 16th percentile, $q=0.16$.

\newpage

\section{Broadening of the $E_1=E_2=0$ peak due to intermediate state scattering}

\begin{figure*}[b]
\centering
\includegraphics[width=16.5cm]{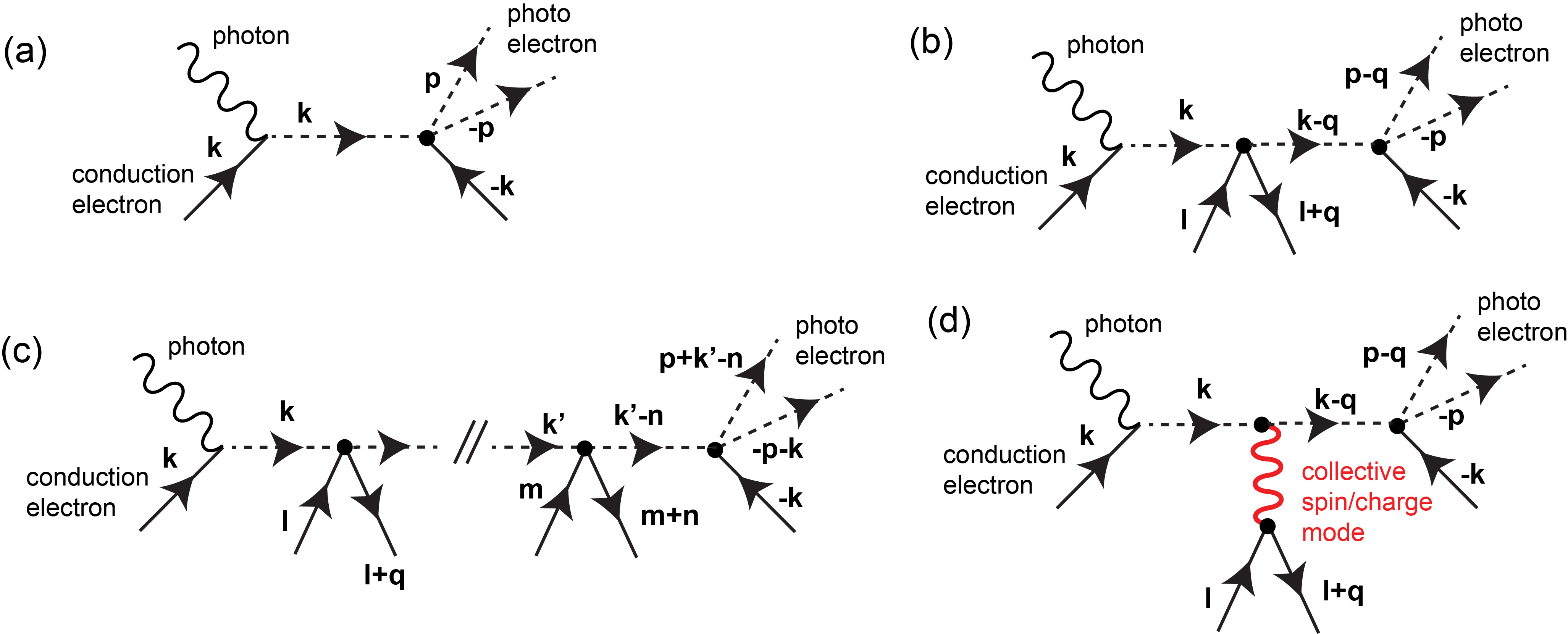}
\caption{(a) Two-step scattering process describing the absorption of a photon and the emission of two photo-electrons. (b) Same process as in (a), but including an intermediate state scattering process in which the photo-electron scatters a conduction electron back into the conduction band. (c) Same as in (a), but with multiple intermediate state scattering processes. (d) Accounting for the strong Coulomb interaction in the conduction band, the intermediate state scattering process can be considered to be mediated by spin or charge collective modes.} 
\label{fig:intermediateState}
\end{figure*}

To understand the broadening in the $(E_1,E_2)$ plane of the experimentally observed peak in the photo-electron counting rate $P^{(2)}$ which is centered at $E_1=E_2=0$, we recall that this peak arises from the emission of two photo-electrons from the same Cooper pair. This contribution to $P^{(2)}$ is referred to as $P^{(2)}_{SC}$ to distinguish it from the contribution, $P^{(2)}_{2cp}$, that is located at higher energies and arises from photo-electrons emitted by two different Cooper pairs. $P^{(2)}_{SC}$ arises from two consecutive processes: the emission of a first photo-electron upon absorption of a photon, and the subsequent scattering (EELS-like) process between the emitted photo-electron and a conduction electron, which leads to the emission of a second photo-electron, as shown in Fig.\ref{fig:intermediateState}(a). For $P^{(2)}_{SC}$, the emitted photo-electrons originate from the same Cooper pair, and thus conduction electrons with momenta ${\bf k}$ and $-{\bf k}$, and thus zero center-of-mass momentum. We assume that the  scattering process between the photo-electron and the conduction electron is mediated by a (screened) Coulomb interaction. This entire process is then described by the Hamiltonian
\begin{align}
    H_{sc} & = \sum_{{\bf k,q},\sigma,\nu} \gamma_{\nu}({\bf q}) d^\dagger_{{\bf k+q},\sigma} c_{{\bf k},\sigma} \left( a_{{\bf q},\nu}  + a^\dagger_{-{\bf q},\nu}\right)  \nonumber \\
    & + \sum_{{\bf k,p,q},\alpha, \beta} V({\bf q}) d^\dagger_{{\bf k}+{\bf q},\alpha} d^\dagger_{{\bf p}-{\bf q},\beta} d_{{\bf p},\beta} c_{{\bf k},\alpha}  + h.c.
\end{align}
Here, $\gamma_{\nu}({\bf q})$ is the effective electron-photon dipole interaction, $d^\dagger_{{\bf k},\sigma} (c_{{\bf k},\sigma})$ creates (destroys) a photo-electron (conduction electron) with momentum ${\bf k}$ and spin $\sigma$, and $V({\bf q}) \sim [{\bf q}^2 + \kappa^{2}]^{-1}$ is the Fourier transform of the (screened) Coulomb interaction, with $\kappa^{-1}$ being the screening length. 

However, the presence of a  Coulomb interaction between the first photo-electron and the conduction electron can not only lead to the emission of a photo-electron, but also to a scattering process in which the conduction electron is scattered back into its band -- a process we refer to as intermediate state scattering --, which is described by the Hamiltonian
\begin{align}
    H_{c} & = \sum_{{\bf k,p,q},\alpha, \beta} V({\bf q}) d^\dagger_{{\bf k}+{\bf q},\alpha} c^\dagger_{{\bf p}-{\bf q},\beta} d_{{\bf k},\beta} c_{{\bf p},\alpha}  + h.c.
\end{align}
Such a process, as shown in Fig.~\ref{fig:intermediateState}(b) transfers momentum, and hence energy, between the photo-electron and the conduction band. As a result, after the emission of the second photo-electron, the two photo-electrons do not possess a zero center-of-mass momentum, and thus now exhibit a finite distribution in the $(E_1,E_2)$ plane. In general, this intermediate state scattering process can occur multiple time before the final emission of second photo-electron, as shown in Fig.~\ref{fig:intermediateState}(c). Moreover, since the emergence of collective spin or charge modes in the cuprate superconductors arises from Coulomb scattering within the conduction band, the inclusion of higher order scattering events in the conduction band allows one to describe the scattering between the photo-electron and conduction band as mediated by a collective spin or charge mode, as shown in Fig.~\ref{fig:intermediateState}(d) (this can be seen, for example, by representing the collective mode as an infinite series of bubble diagrams in the random-phase approximation). This implies that the distribution of the photo-electron energies in the $(E_1,E_2)$ plane is a direct probe for collective modes in the cuprate superconductors. The detailed investigation of how these collective modes affect the intensity distribution of $P^{(2)}_{SC}$ in the $(E_1,E_2)$ plane is reserved for future work. 

\newpage

\section{Broadening of $E_-$: Relative-time dynamics and dephasing}

To construct our model, we use the general approach of \cite{Berakdar1998}. In particular, we start by considering the low-energy electronic states, described by the operator $c_{\bb{k}\sigma,\alpha}$, the intermediate high-energy electronic states, described by $d_{\bb{q}\sigma}$, and the photons, described by $a_{\bb{p}'\gamma}$. Here, the subscript $\alpha$ denotes the set of internal degrees of freedom of the low-energy electrons (like band or orbital), $\sigma$ is the spin, and $\gamma$ is the photon polarization. While the low-energy states are those that form the superconducting condensate below $T_c$, the high-energy electrons are connected to, but are not identified with, the asymptotic photoelectrons with momentum $\bb{p}$ and spin $\sigma$ measured in the detector, as explained in Ref. \cite{Berakdar1998}. The total Hamiltonian can then be written as $H=H_0 + H_1$, with $H_0$ denoting the terms of the Hamiltonian corresponding to the three independent sectors,

\begin{equation}
H_0 = H_\mathrm{ph}+H_\mathrm{low}+H_\mathrm{high},
\label{eq:H0}
\end{equation}
and $H_1$  denoting the terms of the Hamiltonian corresponding to the interactions between the three independent sectors,

\begin{equation}
H_1 = H_{ac}+H_{cd}+H_{acd}\,.
\label{eq:H1_aux}
\end{equation}

We emphasize that $H_0$ does not correspond to a free Hamiltonian, as interactions within each sector are allowed. To proceed, we approximate the photon and high-energy Hamiltonian terms $H_\mathrm{ph}$ and $H_\mathrm{high}$ as free (i.e., they contain only one-body operators) but consider the full interacting $c$-electron Hamiltonian $H_c$, and define $H_\mathrm{low}=H_c$. Importantly, the intermediate $d$-electron modes are distinguished from the asymptotic outgoing photoelectron states. Therefore, even without specifying a microscopic mechanism, we can encode the process by which a $d$-electron with momentum $\bb{p}'$ and spin $\sigma'$ is converted into a photoelectron with momentum $\bb{p}$ and spin $\sigma$ through a matrix element $M_{\sigma \sigma'}(\bb{p},\bb{p}')$.

Our goal is to compute the transition operator for the single-photon-absorption channel, starting from a known coherent photon probe state and the high-energy-electron vacuum at the initial time $t_i$. We project at the final time $t_f$ onto two asymptotic outgoing photoelectrons with momenta $\bb{p}_1$, $\bb{p}_2$ and spins $\sigma_1$, $\sigma_2$. Formally, this transition operator is given by:

\begin{align}
\mathcal{A}_{\sigma_1\sigma_2}^{\mu}(\mathbf p_1,\mathbf p_2;t_f,t_i)
\equiv {}_{\mathrm{out}}\!\langle\mathbf p_1\sigma_1,\mathbf p_2\sigma_2|
\,{}_\gamma\!\langle\mu|U(t_f,t_i)|\mu_0\rangle_{\gamma_{0}}|0\rangle_d .
\label{eq:A}
\end{align}

Here, $|\mu_0\rangle_{\gamma_{0}}$ denotes the known initial coherent photon probe state with polarization $\gamma_0$, $|\mu\rangle_{\gamma}$ denotes the unresolved final photon state with polarization $\gamma$, and $U(t_f,t_i)$ is the time evolution operator of the entire system. The quantity $\mathcal{A}_{\sigma_1\sigma_2}^{\mu}$ is an operator acting on the low-energy electronic sector.

In the experiment, the only known information about the final state is the momenta $\bb{p}_1$ and $\bb{p}_2$ of the photoelectrons detected. In contrast, the photoelectrons spins $\sigma_1$ and $\sigma_2$, the final photon state $\mu$, and the state of the low-energy electrons are not known. Therefore, to obtain the probability density of the photoelectrons momenta, we must trace over these degrees of freedom. For the low-energy electronic sector, this is conveniently expressed in terms of the corresponding density matrix $\rho_c$. Dividing this probability density by the time interval $t_f-t_i$ then gives the two-photoelectron coincidence rate:

\begin{align}
I^{(2)}(\mathbf p_1,\mathbf p_2)\equiv \frac{1}{t_f -t_i}
\sum_{\sigma_1,\sigma_2} \sum_\mu {\rm Tr}_c \left[ \rho_c \mathcal{A}_{\sigma_1 \sigma_2}^{\mu\dagger}
(\mathbf p_1,\mathbf p_2;t_f,t_i) \mathcal{A}_{\sigma_1 \sigma_2}^{\mu} (\mathbf p_1,\mathbf p_2;t_f,t_i) \right] .
\label{eq:I2}
\end{align}

Note that while the expression above is written for a pure coherent initial photon state $|\mu_0\rangle_{\gamma_{0}}$, it is straightforward to generalize it using a general photon density matrix.

To determine $I^{(2)}$, we therefore need to obtain the time evolution operator $U(t_f,t_i)$ of $H=H_0+H_1$, which can be done perturbatively in $H_1$ within the interaction picture. Following previous works \cite{Mahmood2022}, the relevant interactions that account for the two-photoelectron emission by a single photon take place via a two-step process:

\begin{align}
H_{1} = \sum_{\gamma, \bb{k}_i,\sigma_i} \Gamma_{\sigma_1 \sigma_2}^{\gamma}(\bb{k}_1,\bb{k}_2) d_{\bb{k}_1\sigma_1}^{\dagger} c_{\bb{k}_2 \sigma_2}^{\phantom{\dagger}} a_{\gamma}+\sum_{\bb{k}_i,\sigma_i} V\left(\bb{k}_1,\bb{k}_2;\bb{k}_3,\bb{k}_4\right) d^\dagger_{\bb{k_1}\sigma_1} d^\dagger_{\bb{k}_2\sigma_2} d_{\bb{k}_3\sigma_2}^{\phantom{\dagger}}c_{\bb{k}_4\sigma_1}^{\phantom{\dagger}}+ \mathrm{h.c.} \, .
\label{eq:H1}
\end{align}

Here, the first term describes the photoelectron excited by the incoming photon, and $\Gamma_{\sigma \sigma'}^{\gamma}(\bb{p},\bb{k}) $ denotes the vertex for this process. For simplicity, we dropped the additional internal degree of freedom $\alpha$ of the low-energy electrons. Note that the selection rules and momentum conservation are implicitly satisfied by the vertex function. The second term, on the other hand, denotes the EELS-like process by which a high-energy electron interacts with a low-energy electron to simultaneously create two high-energy electrons. The relevant interaction $V\left(\bb{k}_1,\bb{k}_2;\bb{k}_3,\bb{k}_4\right)$ implicitly satisfies momentum conservation and is left unspecified here. Previously, it has been modeled as a screened Coulomb interaction \cite{Mahmood2022}.

For simplicity, we replace the two-step process described in Eq.\ (\ref{eq:H1}) by an effective single-step process. Moreover, since in the experiment a signal is only observed when the system is in the superconducting state, we can restrict our analysis to the case where the two electrons involved in the two-photoelectron emission process form a Cooper pair singlet with zero center-of-mass momentum. We therefore can write:

\begin{align}
H_{1} = H_{1,c} = \frac{1}{2} \sum_{\bb{k},\bb{q}} \mathcal{V}_{\mathrm{eff}}(\bb{k},\bb{q})\, d_{\bb{q}\uparrow}^{\dagger}
d_{-\bb{q} \downarrow}^{\dagger} c_{-\bb{k}\downarrow}^{\phantom{\dagger}} c_{\bb{k} \uparrow}^{\phantom{\dagger}} a_{\gamma}
+ \mathrm{h.c.} 
\label{eq:H1_sudden}
\end{align}

with an effective ``interaction'' $\mathcal V_{\mathrm{eff}}(\bb{k},\bb{q})$. Evaluating the time evolution operator  $U(t_f,t_i)$ perturbatively  is straightforward, yielding:

\begin{equation}
U_I(t_f,t_i) \approx 1 - i\int_{t_i}^{t_f}dt\, H_{1,I}(t) \, ,
\label{eq:U_sudden}
\end{equation}

where the subscript $I$ denotes the interaction picture (we set $\hbar=1$). The final expression for $I^{(2)}(\mathbf p_1,\mathbf p_2)$ can be expressed in terms of the gap function, similarly to Ref. \cite{Berakdar2003,Mahmood2022}. However, the dependence of  $I^{(2)}(\mathbf p_1,\mathbf p_2)$ on the energy difference between the two photoelectrons, $E_- \equiv E_1 - E_2 = \frac{p_1^2}{2m} - \frac{p_2^2}{2m}$, can be inferred without relying on the final expression. The key point is that, in the process described in Eq.\ (\ref{eq:H1_sudden}) (or in \eqref{eq:H1}), the two high-energy electrons are created at the same time. Zero center-of-mass momentum constrains their internal momenta to be opposite, $\bb{q}$ and $-\bb{q}$, but does not require the two detected asymptotic photoelectrons to have equal energies. Because each photoelectron contributes an outgoing Fourier phase, the fact that the two electrons are created at the same time implies that the two Fourier phases combine as:

\begin{equation}
e^{iE_1t}e^{iE_2t} =e^{iE_+t}\, ,
\label{eq:SI-common-time-phase}
\end{equation}

with $E_+=E_1+E_2$ (here, the energy offset associated with the  work function is left implicit). Thus, for the single-photon two-photoelectron process described by Eq.~(\ref{eq:H1_sudden}), the simultaneous production of the pair of high-energy electrons and the absence of subsequent  interactions between them and the material do not introduce an intrinsic relative-time Fourier variable conjugate to the energy difference $E_-$. Of course, due to the momentum dependence in the effective interaction $\mathcal V_{\mathrm{eff}}(\bb{k},\bb{q})$, $I^{(2)}(E_+,E_-)$ can depend on $E_-$ parametrically through matrix elements and kinematic constraints. However, this dependence does not yield an intrinsic broadening of $I^{(2)}(E_+,E_-)$ with respect to $E_-$. Moreover, as explained in the main text, because the observed $E_-$ broadening increases as the temperature is decreased, it cannot be attributed to simple quasi-particle lifetime or instrumental resolution. Instead, within the single-photon two-photoelectron channel considered here, an intrinsic $E_-$ broadening indicates that the high-energy electrons experiences additional interactions in the material following their production, signaling a breakdown of the so-called sudden approximation.  

We thus need to go beyond the sudden approximation and show that $I^{(2)}(E_+,E_-)$ acquires a broadening as a function of $E_-$. Going beyond the sudden approximation requires allowing the high-energy electrons to interact with the material before the photoelectron detection. As discussed in \cite{Berakdar1998}, the high-energy electrons can interact with any degrees of freedom of the material. The second term in Eq.\ (\ref{eq:H1}) is an example of such a process, in which the high-energy electrons interact with the low-energy electrons. To generate an intrinsic relative-time variable in the single-photon two-photoelectron channel considered here, one of the outgoing high-energy electrons must interact with the material after they are generated. To show that the resulting $E_-$ broadening is independent of the microscopic nature of this additional channel, we describe it by an unspecified bosonic field $\Phi$ that couples to the high-energy electronic density through a number-conserving vertex. Therefore, the $H_1$ Hamiltonian acquires a new term:

\begin{align}
H_{1} = H_{1,c} +H_{1,\Phi} = \frac{1}{2} \sum_{\bb{k},\bb{q}} \mathcal{V}_{\mathrm{eff}}(\bb{k},\bb{q})\, d_{\bb{q}\uparrow}^{\dagger}
d_{-\bb{q} \downarrow}^{\dagger} c_{-\bb{k}\downarrow}^{\phantom{\dagger}} c_{\bb{k} \uparrow}^{\phantom{\dagger}} a_{\gamma}
+ \sum_{\sigma,\bb{q},\bb{Q}}
\Lambda(\bb{q},\bb{Q})
 d_{\bb{q}+\bb{Q}\sigma}^{\dagger}
 d_{\bb{q}\sigma}^{\phantom{\dagger}}
\Phi_{\bb{Q}} + \mathrm{h.c.}\, ,
\label{eq:H1_boson}
\end{align}

where $\Lambda(\bb{q},\bb{Q})$ is the relevant coupling constant. In order to evaluate $I^{(2)}$ in Eq.\ (\ref{eq:I2}), we must now also trace over the bosonic degree of freedom. To proceed, we assume that the collective mode is fluctuating, with $\langle \Phi_{\bb{Q}}\rangle_\Phi = 0$ and spectral function defined by:

\begin{align}
\left\langle \Phi_{\bb{Q}_1}^{\dagger}(t') \Phi_{\bb{Q}_2}^{\phantom{\dagger}}(t) \right\rangle_\Phi
=
\delta_{\bb{Q}_1,\bb{Q}_2}
\int_{-\infty}^{\infty}\frac{d\omega}{2\pi}\,
e^{-i\omega(t-t')}S_\Phi(\bb{Q}_2,\omega).
\label{eq:S_Phi}
\end{align}

We can now apply again perturbation theory to obtain the new time evolution operator  $U(t_f,t_i)$ from $H_{1} = H_{1,c} +H_{1,\Phi}$. While several terms appear, the leading term describing a later scattering event involving one of the two high-energy electrons is the second-order cross term:

\begin{equation}
(-i)^2 \int_{t_i}^{t_f}dt \int_t^{t_f}dt'\,
H_{1,\Phi,I}(t')H_{1,c,I}(t).
\label{eq:Dyson-one-Phi-ordering-2e}
\end{equation}

The reverse ordering term $H_{1,c,I}(t')H_{1,\Phi,I}(t)$ does not contribute in the initial $d$-electron vacuum because $H_{1,\Phi}$ conserves the number of high-energy electrons and therefore cannot act before $H_{1,c}$ creates the two high-energy electrons. Inserting this term in the expression of the transition operator $\mathcal{A}_{\sigma_1\sigma_2}^{\mu}(\mathbf p_1,\mathbf p_2;t_f,t_i)$ in Eq.\ (\ref{eq:A}) generates a transition amplitude that is linear in $\Phi$ and of order $\mathcal{V}_{\mathrm{eff}}\Lambda$. Since $\langle\Phi\rangle_\Phi=0$, its interference with the amplitude containing no $H_{1,\Phi}$ term vanishes. The leading non-zero contribution is therefore the product of the linear-in-$\Phi$ amplitude and its Hermitian conjugate, which is of order $\mathcal{V}_{\mathrm{eff}}^2\Lambda^2$ and given by:

\begin{align}
\tilde{I}^{(2)}(E_+,E_-)= \sum_{\bb{Q}} S_\Phi(\bb{Q},E_+-\Omega)
\left|\sum_{\bb{k},\bb{q}} F_{\bb{k}} \,
\mathcal{C}(\bb{k},\bb{q};\bb{Q},\mathbf p_1,\mathbf p_2) \,
\mathcal{D}_{\bb{q}}(E_+,E_-;E_+ - \Omega)\right|^2 \, .
\label{eq:I_tilde}
\end{align}

Here, $\tilde I^{(2)}$ is the positive contribution with one explicit $\Phi$ vertex in the amplitude and one in its conjugate. We note that this is not the complete coincidence rate at order $\Lambda^2$, which also contains the interference of the no-$\Phi$ amplitude with amplitudes containing two explicit $\Phi$ vertices. In Eq.\ (\ref{eq:I_tilde}), $\Omega$ is the incoming photon frequency, $\mathcal{C}(\bb{k},\bb{q};\bb{Q},\mathbf p_1,\mathbf p_2)$ is a form factor combining the interaction vertices and the outgoing-state overlaps, and $F_{\bb{k}}$ is the equal-time anomalous pair amplitude (computed here within mean field theory),

\begin{equation}
F_{\bb{k}}\equiv
\left\langle c_{-\bb{k}\downarrow}c_{\bb{k}\uparrow}\right\rangle_c
= \frac{\Delta_{\bb{k}}}{2\sqrt{\xi_{\bb{k}}^2+|\Delta_{\bb{k}}|^2}} \,\tanh\left(\frac{\beta \sqrt{\xi_{\bb{k}}^2+|\Delta_{\bb{k}}|^2}}{2}\right)
\label{eq:F}
\end{equation}
where $\Delta_{\bb{k}}$ is the gap function and $\xi_{\bb{k}}$ is the normal-state dispersion of the low-energy electrons. Under the additional assumption that the matrix elements and high-energy-electron propagators for the two scattering processes are approximately equal in the measured kinematic window, the scattering kernel is

\begin{align}
\mathcal{D}_{\bb{q}}(E_+,E_-;\omega) = \int_0^\infty d\tau\,
 e^{-\Gamma_d\tau/2} e^{+i(E_+/2-\omega-E_{\bb{q}})\tau}
\cos\left(\frac{E_-\tau}{2}\right).
\label{eq:D}
\end{align}
where $E_{\bb{q}}$ is the energy of the high-energy electrons offset by the work function. In the equation above, $\Gamma_d$ is a phenomenological effective damping/dephasing rate of the high-energy-electron propagator, which may receive contributions both from scattering off of the $\Phi$ mode and from other unspecified interactions (these contributions are not computed separately here). The central result underlying the $E_-$ broadening of $I^{(2)}$ is encoded in the function $\mathcal{D}_{\bb{q}}(E_+,E_-;\omega)$ defined above. It arises because, due to the exchange of energy between the high-energy electrons and the collective bosonic modes, there are two time scales involved. Specifically, at time $t$, the two high-energy electrons are created and at a later time $t+\tau$, one of them interacts with the collective mode. The outgoing Fourier phases for each scattering process multiply, whereas the two amplitudes corresponding to delayed scattering from either outgoing electron are added coherently. The latter can be approximated as:

\begin{align}
\ee^{iE_1\tau}+\ee^{iE_2\tau} &= 2\ee^{iE_+\tau/2} \cos\left(\frac{E_-\tau}{2}\right),
\label{eq:SI-branch-interference}
\end{align}
which is the same expression that appears in the integrand of Eq.\ (\ref{eq:D}). The emergence of an $E_-$ broadening can be directly seen by evaluating the integral in Eq.\ (\ref{eq:D}) at the resonance condition

\begin{equation}
\omega_{\mathrm{res}}=\frac{E_+}{2}-E_{\bb{q}},
\end{equation}
which, using $\omega=E_+-\Omega$ in Eq.\ (\ref{eq:I_tilde}), corresponds to a resonance condition for the energy of the $d$ electrons,  $2E_{\bb{q}}= 2\Omega-E_+$. At resonance,

\begin{equation}
\left|\mathcal{D}_{\bb{q}}(E_+,E_-;\omega_{\mathrm{res}})\right|^2
= \frac{4\Gamma_d^2} {\left(\Gamma_d^2+E_-^2\right)^2}.
\label{eq:SI-Eminus-lineshape}
\end{equation}

Thus, the broadening is controlled by the effective high-energy-electron damping/dephasing rate $\Gamma_d$, as well as by the spectral weight of the collective mode through $S_\Phi$, as shown by Eq.~(\ref{eq:I_tilde}). A broadening that increases as the temperature decreases is therefore indicative of enhanced damping/dephasing of the outgoing high-energy electrons and/or a temperature-dependent reweighting of the contributing scattering processes. It does not, by itself, necessarily establish that the collective mode becomes softer as the temperature is lowered, since such a conclusion would require independent information about the temperature dependence of $S_\Phi$, which is beyond the scope of this work.

%% file: combined.bbl
\begin{thebibliography}{10}
\expandafter\ifx\csname url\endcsname\relax
  \def\url#1{\texttt{#1}}\fi
\expandafter\ifx\csname urlprefix\endcsname\relax\def\urlprefix{URL }\fi
\providecommand{\bibinfo}[2]{#2}
\providecommand{\eprint}[2][]{\url{#2}}

\bibitem{BCS}
\bibinfo{author}{Bardeen, J.}, \bibinfo{author}{Cooper, L.~N.} \& \bibinfo{author}{Schrieffer, J.~R.}
\newblock \bibinfo{title}{Theory of superconductivity}.
\newblock \emph{\bibinfo{journal}{Phys. Rev.}} \textbf{\bibinfo{volume}{108}}, \bibinfo{pages}{1175--1204} (\bibinfo{year}{1957}).

\bibitem{Tinkham2004}
\bibinfo{author}{Tinkham, M.}
\newblock \emph{\bibinfo{title}{Introduction to Superconductivity: v. 1}}.
\newblock Dover Books on Physics (\bibinfo{publisher}{Dover Publications}, \bibinfo{address}{Mineola, NY}, \bibinfo{year}{2004}).

\bibitem{Damascelli_2003}
\bibinfo{author}{Damascelli, A.}, \bibinfo{author}{Hussain, Z.} \& \bibinfo{author}{Shen, Z.-X.}
\newblock \bibinfo{title}{Angle-resolved photoemission studies of the cuprate superconductors}.
\newblock \emph{\bibinfo{journal}{Rev. Mod. Phys.}} \textbf{\bibinfo{volume}{75}}, \bibinfo{pages}{473--541} (\bibinfo{year}{2003}).

\bibitem{Lee2006}
\bibinfo{author}{Lee, P.~A.}, \bibinfo{author}{Nagaosa, N.} \& \bibinfo{author}{Wen, X.-G.}
\newblock \bibinfo{title}{Doping a {M}ott insulator: Physics of high-temperature superconductivity}.
\newblock \emph{\bibinfo{journal}{Reviews of Modern Physics}} \textbf{\bibinfo{volume}{78}}, \bibinfo{pages}{17--85} (\bibinfo{year}{2006}).

\bibitem{Norman2011}
\bibinfo{author}{Norman, M.~R.}
\newblock \bibinfo{title}{The challenge of unconventional superconductivity}.
\newblock \emph{\bibinfo{journal}{Science}} \textbf{\bibinfo{volume}{332}}, \bibinfo{pages}{196–200} (\bibinfo{year}{2011}).

\bibitem{Sobota2021}
\bibinfo{author}{Sobota, J.~A.}, \bibinfo{author}{He, Y.} \& \bibinfo{author}{Shen, Z.-X.}
\newblock \bibinfo{title}{Angle-resolved photoemission studies of quantum materials}.
\newblock \emph{\bibinfo{journal}{Reviews of Modern Physics}} \textbf{\bibinfo{volume}{93}} (\bibinfo{year}{2021}).

\bibitem{Keimer2015}
\bibinfo{author}{Keimer, B.}, \bibinfo{author}{Kivelson, S.~A.}, \bibinfo{author}{Norman, M.~R.}, \bibinfo{author}{Uchida, S.} \& \bibinfo{author}{Zaanen, J.}
\newblock \bibinfo{title}{From quantum matter to high-temperature superconductivity in copper oxides}.
\newblock \emph{\bibinfo{journal}{Nature}} \textbf{\bibinfo{volume}{518}}, \bibinfo{pages}{179–186} (\bibinfo{year}{2015}).

\bibitem{Fradkin2015}
\bibinfo{author}{Fradkin, E.}, \bibinfo{author}{Kivelson, S.~A.} \& \bibinfo{author}{Tranquada, J.~M.}
\newblock \bibinfo{title}{Colloquium: Theory of intertwined orders in high temperature superconductors}.
\newblock \emph{\bibinfo{journal}{Reviews of Modern Physics}} \textbf{\bibinfo{volume}{87}}, \bibinfo{pages}{457–482} (\bibinfo{year}{2015}).

\bibitem{Abanov2003}
\bibinfo{author}{Abanov, A.}, \bibinfo{author}{Chubukov, A.~V.} \& \bibinfo{author}{Schmalian, J.}
\newblock \bibinfo{title}{Quantum-critical theory of the spin-fermion model and its application to cuprates: Normal state analysis}.
\newblock \emph{\bibinfo{journal}{Advances in Physics}} \textbf{\bibinfo{volume}{52}}, \bibinfo{pages}{119--218} (\bibinfo{year}{2003}).

\bibitem{Daix2026}
\bibinfo{author}{Daix, C.} \emph{et~al.}
\newblock \bibinfo{title}{Observing spatial charge and spin correlations in a strongly interacting fermi gas}.
\newblock \emph{\bibinfo{journal}{Phys. Rev. Lett.}} \textbf{\bibinfo{volume}{136}}, \bibinfo{pages}{153402} (\bibinfo{year}{2026}).

\bibitem{Haak1978}
\bibinfo{author}{Haak, H.~W.}, \bibinfo{author}{Sawatzky, G.~A.} \& \bibinfo{author}{Thomas, T.~D.}
\newblock \bibinfo{title}{Auger-{{Photoelectron Coincidence Measurements}} in {{Copper}}}.
\newblock \emph{\bibinfo{journal}{Phys. Rev. Lett.}} \textbf{\bibinfo{volume}{41}}, \bibinfo{pages}{1825--1827} (\bibinfo{year}{1978}).

\bibitem{Herrmann1998}
\bibinfo{author}{Herrmann, R.}, \bibinfo{author}{Samarin, S.}, \bibinfo{author}{Schwabe, H.} \& \bibinfo{author}{Kirschner, J.}
\newblock \bibinfo{title}{Two electron photoemission in solids}.
\newblock \emph{\bibinfo{journal}{Phys. Rev. Lett.}} \textbf{\bibinfo{volume}{81}}, \bibinfo{pages}{2148--2151} (\bibinfo{year}{1998}).

\bibitem{Schumann2007}
\bibinfo{author}{Schumann, F.~O.}, \bibinfo{author}{Winkler, C.} \& \bibinfo{author}{Kirschner, J.}
\newblock \bibinfo{title}{Correlation effects in two electron photoemission}.
\newblock \emph{\bibinfo{journal}{Phys. Rev. Lett.}} \textbf{\bibinfo{volume}{98}}, \bibinfo{pages}{257604} (\bibinfo{year}{2007}).

\bibitem{Hattass2008}
\bibinfo{author}{Hattass, M.} \emph{et~al.}
\newblock \bibinfo{title}{Dynamics of two-electron photoemission from {{Cu}}(111)}.
\newblock \emph{\bibinfo{journal}{Phys. Rev. B}} \textbf{\bibinfo{volume}{77}}, \bibinfo{pages}{165432} (\bibinfo{year}{2008}).

\bibitem{Schumann2011}
\bibinfo{author}{Schumann, F.~O.}, \bibinfo{author}{Dhaka, R.~S.}, \bibinfo{author}{{van Riessen}, G.~A.}, \bibinfo{author}{Wei, Z.} \& \bibinfo{author}{Kirschner, J.}
\newblock \bibinfo{title}{Surface state and resonance effects in electron-pair emission from {{Cu}}(111)}.
\newblock \emph{\bibinfo{journal}{Phys. Rev. B}} \textbf{\bibinfo{volume}{84}}, \bibinfo{pages}{125106} (\bibinfo{year}{2011}).

\bibitem{Schumann2012}
\bibinfo{author}{Schumann, F.~O.} \emph{et~al.}
\newblock \bibinfo{title}{Electron pair emission from a highly correlated material}.
\newblock \emph{\bibinfo{journal}{Phys. Rev. B}} \textbf{\bibinfo{volume}{86}}, \bibinfo{pages}{035131} (\bibinfo{year}{2012}).

\bibitem{Chiang2020}
\bibinfo{author}{Chiang, C.-T.} \emph{et~al.}
\newblock \bibinfo{title}{Laser-based double photoemission spectroscopy at surfaces}.
\newblock \emph{\bibinfo{journal}{Progress in Surface Science}} \textbf{\bibinfo{volume}{95}}, \bibinfo{pages}{100572} (\bibinfo{year}{2020}).

\bibitem{Schumann2020}
\bibinfo{author}{Schumann, F.~O.}, \bibinfo{author}{Kirschner, J.} \& \bibinfo{author}{Berakdar, J.}
\newblock \bibinfo{title}{Imaging momentum–space two‐particle correlations at surfaces}.
\newblock \emph{\bibinfo{journal}{physica status solidi (b)}} \textbf{\bibinfo{volume}{257}} (\bibinfo{year}{2020}).

\bibitem{Berakdar1998}
\bibinfo{author}{Berakdar, J.}
\newblock \bibinfo{title}{Emission of correlated electron pairs following single-photon absorption by solids and surfaces}.
\newblock \emph{\bibinfo{journal}{Phys. Rev. B}} \textbf{\bibinfo{volume}{58}}, \bibinfo{pages}{9808–9816} (\bibinfo{year}{1998}).

\bibitem{Berakdar2003}
\bibinfo{author}{Kouzakov, K.~A.} \& \bibinfo{author}{Berakdar, J.}
\newblock \bibinfo{title}{Photoinduced emission of {C}ooper pairs from superconductors}.
\newblock \emph{\bibinfo{journal}{Phys. Rev. Lett.}} \textbf{\bibinfo{volume}{91}}, \bibinfo{pages}{257007} (\bibinfo{year}{2003}).

\bibitem{KOUZAKOV2007}
\bibinfo{author}{Kouzakov, K.} \& \bibinfo{author}{Berakdar, J.}
\newblock \bibinfo{title}{Mechanisms of superconductivity studied by two-particle emission}.
\newblock \emph{\bibinfo{journal}{Journal of Electron Spectroscopy and Related Phenomena}} \textbf{\bibinfo{volume}{161}}, \bibinfo{pages}{121--124} (\bibinfo{year}{2007}).
\newblock \bibinfo{note}{Many particle spectroscopy of atoms, molecules, clusters and surfaces}.

\bibitem{Wong2024}
\bibinfo{author}{Wong, K.~H.}, \bibinfo{author}{Patwardhan, A.}, \bibinfo{author}{Abbamonte, P.}, \bibinfo{author}{Mahmood, F.} \& \bibinfo{author}{Morr, D.~K.}
\newblock \bibinfo{title}{Two-electron photoemission spectroscopy in topological superconductors}.
\newblock \emph{\bibinfo{journal}{Phys. Rev. B}} \textbf{\bibinfo{volume}{109}} (\bibinfo{year}{2024}).

\bibitem{Wong2025}
\bibinfo{author}{Wong, K.~H.} \emph{et~al.}
\newblock \bibinfo{title}{Measuring the {D}ebye energy in superconductors via two-electron photoemission spectroscopy}.
\newblock \emph{\bibinfo{journal}{Phys. Rev. B}} \textbf{\bibinfo{volume}{112}}, \bibinfo{pages}{134523} (\bibinfo{year}{2025}).

\bibitem{Su_2024}
\bibinfo{author}{Su, Y.}, \bibinfo{author}{Cao, K.} \& \bibinfo{author}{Zhang, C.}
\newblock \bibinfo{title}{Coincidence detection probability of photoemission measurement}.
\newblock \emph{\bibinfo{journal}{Journal of Physics: Condensed Matter}} \textbf{\bibinfo{volume}{36}}, \bibinfo{pages}{505602} (\bibinfo{year}{2024}).

\bibitem{Su2020}
\bibinfo{author}{Su, Y.} \& \bibinfo{author}{Zhang, C.}
\newblock \bibinfo{title}{Coincidence angle-resolved photoemission spectroscopy: Proposal for detection of two-particle correlations}.
\newblock \emph{\bibinfo{journal}{Phys. Rev. B}} \textbf{\bibinfo{volume}{101}}, \bibinfo{pages}{205110} (\bibinfo{year}{2020}).

\bibitem{Devereaux2023}
\bibinfo{author}{Devereaux, T.~P.} \emph{et~al.}
\newblock \bibinfo{title}{Angle-resolved pair photoemission theory for correlated electrons}.
\newblock \emph{\bibinfo{journal}{Phys. Rev. B}} \textbf{\bibinfo{volume}{108}}, \bibinfo{pages}{165134} (\bibinfo{year}{2023}).

\bibitem{Kemper2025}
\bibinfo{author}{Kemper, A.~F.} \emph{et~al.}
\newblock \bibinfo{title}{Observing two-electron interactions with correlation-arpes}.
\newblock \emph{\bibinfo{journal}{Phys. Rev. B}} \textbf{\bibinfo{volume}{112}}, \bibinfo{pages}{035168} (\bibinfo{year}{2025}).

\bibitem{Bonca2026}
\bibinfo{author}{Bonca, J.}, \bibinfo{author}{Damascelli, A.} \& \bibinfo{author}{Berciu, M.}
\newblock \bibinfo{title}{Fingerprints of preformed pairs in two-electron angle-resolved photoemission spectroscopy}.
\newblock \emph{\bibinfo{journal}{Phys. Rev. Lett.}} \textbf{\bibinfo{volume}{136}}, \bibinfo{pages}{196504} (\bibinfo{year}{2026}).

\bibitem{Setup}
\bibinfo{author}{Zwettler, J.} \emph{et~al.}
\newblock \bibinfo{title}{An extreme ultraviolet 2e-{ARPES} setup based on dual time-of-flight analyzers}.
\newblock \emph{\bibinfo{journal}{Journal of Electron Spectroscopy and Related Phenomena}} \textbf{\bibinfo{volume}{270}}, \bibinfo{pages}{147417} (\bibinfo{year}{2024}).

\bibitem{Ding1996}
\bibinfo{author}{Ding, H.} \emph{et~al.}
\newblock \bibinfo{title}{Angle-resolved photoemission spectroscopy study of the superconducting gap anisotropy in {B}i$_2${S}r$_2${C}a{C}u$2${O}${8+x}$}.
\newblock \emph{\bibinfo{journal}{Phys. Rev. B}} \textbf{\bibinfo{volume}{54}}, \bibinfo{pages}{R9678–R9681} (\bibinfo{year}{1996}).

\bibitem{Hashimoto2014}
\bibinfo{author}{Hashimoto, M.}, \bibinfo{author}{Vishik, I.~M.}, \bibinfo{author}{He, R.-H.}, \bibinfo{author}{Devereaux, T.~P.} \& \bibinfo{author}{Shen, Z.-X.}
\newblock \bibinfo{title}{Energy gaps in high-transition-temperature cuprate superconductors}.
\newblock \emph{\bibinfo{journal}{Nature Physics}} \textbf{\bibinfo{volume}{10}}, \bibinfo{pages}{483–495} (\bibinfo{year}{2014}).

\bibitem{Kostanovskiy2015}
\bibinfo{author}{Kostanovskiy, I.}, \bibinfo{author}{Schumann, F.~O.}, \bibinfo{author}{Aliaev, Y.}, \bibinfo{author}{Wei, Z.} \& \bibinfo{author}{Kirschner, J.}
\newblock \bibinfo{title}{Core-resonant double photoemission from palladium films}.
\newblock \emph{\bibinfo{journal}{Journal of Physics: Condensed Matter}} \textbf{\bibinfo{volume}{28}}, \bibinfo{pages}{015601} (\bibinfo{year}{2015}).

\bibitem{Mahmood2022}
\bibinfo{author}{Mahmood, F.}, \bibinfo{author}{Devereaux, T.}, \bibinfo{author}{Abbamonte, P.} \& \bibinfo{author}{Morr, D.~K.}
\newblock \bibinfo{title}{Distinguishing finite-momentum superconducting pairing states with two-electron photoemission spectroscopy}.
\newblock \emph{\bibinfo{journal}{Phys. Rev. B}} \textbf{\bibinfo{volume}{105}}, \bibinfo{pages}{064515} (\bibinfo{year}{2022}).

\bibitem{Agterberg2020}
\bibinfo{author}{Agterberg, D.~F.} \emph{et~al.}
\newblock \bibinfo{title}{The physics of pair-density waves: cuprate superconductors and beyond}.
\newblock \emph{\bibinfo{journal}{Annual Review of Condensed Matter Physics}} \textbf{\bibinfo{volume}{11}}, \bibinfo{pages}{231--270} (\bibinfo{year}{2020}).

\bibitem{Fong1999}
\bibinfo{author}{Fong, H.~F.} \emph{et~al.}
\newblock \bibinfo{title}{Neutron scattering from magnetic excitations in {B}i$_2${S}r$_2${C}a{C}u$2${O}${8+\delta}$}.
\newblock \emph{\bibinfo{journal}{Nature}} \textbf{\bibinfo{volume}{398}}, \bibinfo{pages}{588–591} (\bibinfo{year}{1999}).

\bibitem{Guarise2014}
\bibinfo{author}{Guarise, M.} \emph{et~al.}
\newblock \bibinfo{title}{Anisotropic softening of magnetic excitations along the nodal direction in superconducting cuprates}.
\newblock \emph{\bibinfo{journal}{Nature Communications}} \textbf{\bibinfo{volume}{5}} (\bibinfo{year}{2014}).

\bibitem{Mitrano2018}
\bibinfo{author}{Mitrano, M.} \emph{et~al.}
\newblock \bibinfo{title}{Anomalous density fluctuations in a strange metal}.
\newblock \emph{\bibinfo{journal}{Proceedings of the National Academy of Sciences}} \textbf{\bibinfo{volume}{115}}, \bibinfo{pages}{5392–5396} (\bibinfo{year}{2018}).

\bibitem{Emery1995}
\bibinfo{author}{Emery, V.~J.} \& \bibinfo{author}{Kivelson, S.~A.}
\newblock \bibinfo{title}{Importance of phase fluctuations in superconductors with small superfluid density}.
\newblock \emph{\bibinfo{journal}{Nature}} \textbf{\bibinfo{volume}{374}}, \bibinfo{pages}{434–437} (\bibinfo{year}{1995}).

\bibitem{Corson1999}
\bibinfo{author}{Corson, J.}, \bibinfo{author}{Mallozzi, R.}, \bibinfo{author}{Orenstein, J.}, \bibinfo{author}{Eckstein, J.~N.} \& \bibinfo{author}{Bozovic, I.}
\newblock \bibinfo{title}{Vanishing of phase coherence in underdoped {B}i$_2${S}r$_2${C}a{C}u$2${O}${8+\delta}$}.
\newblock \emph{\bibinfo{journal}{Nature}} \textbf{\bibinfo{volume}{398}}, \bibinfo{pages}{221–223} (\bibinfo{year}{1999}).

\bibitem{Corson_2000}
\bibinfo{author}{Corson, J.}, \bibinfo{author}{Orenstein, J.}, \bibinfo{author}{Oh, S.}, \bibinfo{author}{O'Donnell, J.} \& \bibinfo{author}{Eckstein, J.~N.}
\newblock \bibinfo{title}{Nodal quasiparticle lifetime in the superconducting state of {B}i$_2${S}r$_2${C}a{C}u$2${O}${8+\delta}$}.
\newblock \emph{\bibinfo{journal}{Phys. Rev. Lett.}} \textbf{\bibinfo{volume}{85}}, \bibinfo{pages}{2569--2572} (\bibinfo{year}{2000}).

\bibitem{Pan2001}
\bibinfo{author}{Pan, S.~H.} \emph{et~al.}
\newblock \bibinfo{title}{Microscopic electronic inhomogeneity in the high-tc superconductor {B}i$_2${S}r$_2${C}a{C}u$2${O}${8+\delta}$}.
\newblock \emph{\bibinfo{journal}{Nature}} \textbf{\bibinfo{volume}{413}}, \bibinfo{pages}{282–285} (\bibinfo{year}{2001}).

\bibitem{McElroy2005}
\bibinfo{author}{McElroy, K.} \emph{et~al.}
\newblock \bibinfo{title}{Atomic-scale sources and mechanism of nanoscale electronic disorder in {B}i$_2${S}r$_2${C}a{C}u$2${O}${8+\delta}$}.
\newblock \emph{\bibinfo{journal}{Science}} \textbf{\bibinfo{volume}{309}}, \bibinfo{pages}{1048–1052} (\bibinfo{year}{2005}).

\bibitem{Pelc2018}
\bibinfo{author}{Pelc, D.} \emph{et~al.}
\newblock \bibinfo{title}{Emergence of superconductivity in the cuprates via a universal percolation process}.
\newblock \emph{\bibinfo{journal}{Nature Communications}} \textbf{\bibinfo{volume}{9}} (\bibinfo{year}{2018}).

\bibitem{Bootstrap_intro_new}
\bibinfo{author}{Chernick, M.~R.}
\newblock \emph{\bibinfo{title}{Bootstrap methods : a guide for practitioners and researchers}}.
\newblock Wiley series in probability and statistics (\bibinfo{publisher}{Wiley-Interscience}, \bibinfo{address}{Hoboken, N.J}), \bibinfo{edition}{2nd ed.} edn.

\end{thebibliography}


\begin{thebibliography}{10}
\expandafter\ifx\csname url\endcsname\relax
  \def\url#1{\texttt{#1}}\fi
\expandafter\ifx\csname urlprefix\endcsname\relax\def\urlprefix{URL }\fi
\providecommand{\bibinfo}[2]{#2}
\providecommand{\eprint}[2][]{\url{#2}}

\bibitem{Mahmood2022}
\bibinfo{author}{Mahmood, F.}, \bibinfo{author}{Devereaux, T.}, \bibinfo{author}{Abbamonte, P.} \& \bibinfo{author}{Morr, D.~K.}
\newblock \bibinfo{title}{Distinguishing finite-momentum superconducting pairing states with two-electron photoemission spectroscopy}.
\newblock \emph{\bibinfo{journal}{Phys. Rev. B}} \textbf{\bibinfo{volume}{105}}, \bibinfo{pages}{064515} (\bibinfo{year}{2022}).

\bibitem{Bi2212_gap_sign}
\bibinfo{author}{Gao, Q.} \emph{et~al.}
\newblock \bibinfo{title}{{ARPES} detection of superconducting gap sign in unconventional superconductors}.
\newblock \emph{\bibinfo{journal}{Nature Communications}} \textbf{\bibinfo{volume}{15}}, \bibinfo{pages}{4538--9} (\bibinfo{year}{2024}).

\bibitem{Bi2212_gap_sign_2}
\bibinfo{author}{Gao, Q.} \emph{et~al.}
\newblock \bibinfo{title}{Selective hybridization between the main band and the superstructure band in the {B}i$_{2}${Sr}$_{2}${C}a{C}u$_{2}${O}$_{8+\ensuremath{\delta}}$ superconductor}.
\newblock \emph{\bibinfo{journal}{Phys. Rev. B}} \textbf{\bibinfo{volume}{101}}, \bibinfo{pages}{014513} (\bibinfo{year}{2020}).

\bibitem{Bi2212_ARPES_1}
\bibinfo{author}{Kordyuk, A.~A.} \emph{et~al.}
\newblock \bibinfo{title}{Origin of the peak-dip-hump line shape in the superconducting-state $(\ensuremath{\pi},0)$ photoemission spectra of {B}i$_{2}${Sr}$_{2}${C}a{C}u$_{2}${O}$_{8+\ensuremath{\delta}}$}.
\newblock \emph{\bibinfo{journal}{Phys. Rev. Lett.}} \textbf{\bibinfo{volume}{89}}, \bibinfo{pages}{077003} (\bibinfo{year}{2002}).

\bibitem{Dynes}
\bibinfo{author}{Dynes, R.~C.}, \bibinfo{author}{Narayanamurti, V.} \& \bibinfo{author}{Garno, J.~P.}
\newblock \bibinfo{title}{Direct measurement of quasiparticle-lifetime broadening in a strong-coupled superconductor}.
\newblock \emph{\bibinfo{journal}{Phys. Rev. Lett.}} \textbf{\bibinfo{volume}{41}}, \bibinfo{pages}{1509--1512} (\bibinfo{year}{1978}).

\bibitem{Bi2212_gap_1}
\bibinfo{author}{Sun, X.} \emph{et~al.}
\newblock \bibinfo{title}{Temperature evolution of energy gap and band structure in the superconducting and pseudogap states of {B}i$_{2}${Sr}$_{2}${C}a{C}u$_{2}${O}$_{8+\ensuremath{\delta}}$ superconductor revealed by laser-based angle-resolved photoemission spectroscopy}.
\newblock \emph{\bibinfo{journal}{Chinese Physics Letters}} \textbf{\bibinfo{volume}{35}}, \bibinfo{pages}{017401} (\bibinfo{year}{2018}).

\bibitem{Bi2212_gap_2}
\bibinfo{author}{Vishik, I.~M.} \emph{et~al.}
\newblock \bibinfo{title}{Doping-dependent nodal {F}ermi velocity of the high-temperature superconductor {B}i$_{2}${Sr}$_{2}${C}a{C}u$_{2}${O}$_{8+\ensuremath{\delta}}$ revealed using high-resolution angle-resolved photoemission spectroscopy}.
\newblock \emph{\bibinfo{journal}{Phys. Rev. Lett.}} \textbf{\bibinfo{volume}{104}}, \bibinfo{pages}{207002} (\bibinfo{year}{2010}).

\bibitem{Leitner2021}
\bibinfo{author}{Leitner, T.} \emph{et~al.}
\newblock \bibinfo{title}{The {CoESCA} station at {BESSY}: {A}uger electron–photoelectron coincidences from surfaces demonstrated for {Ag MNN}}.
\newblock \emph{\bibinfo{journal}{Journal of Electron Spectroscopy and Related Phenomena}} \textbf{\bibinfo{volume}{250}}, \bibinfo{pages}{147075} (\bibinfo{year}{2021}).

\bibitem{Bootstrap_intro_new}
\bibinfo{author}{Chernick, M.~R.}
\newblock \emph{\bibinfo{title}{Bootstrap methods : a guide for practitioners and researchers}}.
\newblock Wiley series in probability and statistics (\bibinfo{publisher}{Wiley-Interscience}, \bibinfo{address}{Hoboken, N.J}), \bibinfo{edition}{2nd ed.} edn.

\bibitem{Quantile}
\bibinfo{author}{Hyndman, R.~J.} \& \bibinfo{author}{Fan, Y.}
\newblock \bibinfo{title}{Sample quantiles in statistical packages}.
\newblock \emph{\bibinfo{journal}{The American Statistician}} \textbf{\bibinfo{volume}{50}}, \bibinfo{pages}{361--365} (\bibinfo{year}{1996}).

\bibitem{Berakdar1998}
\bibinfo{author}{Berakdar, J.}
\newblock \bibinfo{title}{Emission of correlated electron pairs following single-photon absorption by solids and surfaces}.
\newblock \emph{\bibinfo{journal}{Phys. Rev. B}} \textbf{\bibinfo{volume}{58}}, \bibinfo{pages}{9808–9816} (\bibinfo{year}{1998}).

\bibitem{Berakdar2003}
\bibinfo{author}{Kouzakov, K.~A.} \& \bibinfo{author}{Berakdar, J.}
\newblock \bibinfo{title}{Photoinduced emission of {C}ooper pairs from superconductors}.
\newblock \emph{\bibinfo{journal}{Phys. Rev. Lett.}} \textbf{\bibinfo{volume}{91}}, \bibinfo{pages}{257007} (\bibinfo{year}{2003}).

\end{thebibliography}
